\documentclass[twocolumn,aps,prl,showpacs]{revtex4-1}

\usepackage{times}
\usepackage{graphicx}
\usepackage{amssymb}
\usepackage{amsmath}

\begin{document} 

\title{Local observation of pair-condensation in a Fermi gas at unitarity} 

\author{M. G. Lingham, K. Fenech, S. Hoinka, and C. J. Vale$^{\ast}$} 

\affiliation{Centre for Atom Optics and Ultrafast Spectroscopy, Swinburne University of Technology, Melbourne 3122, Australia. \\{$^\ast$To whom correspondence should be addressed; E-mail:  cvale@swin.edu.au} }

\begin{abstract}

We present measurements of the local (homogeneous) density-density response function of a Fermi gas at unitarity using spatially resolved Bragg spectroscopy.  By analyzing the Bragg response across one axis of the cloud we extract the response function for a uniform gas which shows a clear signature of the Bose-Einstein condensation of pairs of fermions when the local temperature drops below the superfluid transition temperature.  The method we use for local measurement generalizes a scheme for obtaining the local pressure in a harmonically trapped cloud from the line density and can be adapted to provide any homogeneous parameter satisfying the local density approximation.

\end{abstract}

\pacs{03.75.Hh, 03.75.Ss, 05.30.Fk}

\date{\today}

\maketitle 

Phase transitions and critical phenomena are central topics in low temperature physics in settings ranging from the solid state \cite{Sachdev11qpt} to superfluids \cite{Griffin93eia} and cold atomic gases \cite{Bloch08mbp}.  Clear identification of phase boundaries, however, can prove challenging in experiments.  A key example is a Fermi gas with resonant interactions where bulk superfluidity was definitively shown via the observation of vortex lattices \cite{Zwierlein05vas}, yet detailed characterizations of the phase transition and superfluid fraction have taken much longer \cite{Kinast05hca,Nascimbene10ett,Ku12rts,Sidorenkov13ssa}.  Superfluidity in three-dimensional (3D) Fermi gases occurs simultaneously with the formation of a Bose-Einstein condensate of fermion pairs.  In a spin-balanced Fermi gas at unitarity, this pair condensation is difficult to observe directly as it leads to only a very slight change in the atomic density \cite{Ketterle08mpa}.  Nonetheless, condensation has been verified using rapid sweeps of the effective attractive interaction during time-of-flight expansion, in which pairs are transformed into tightly bound molecules that preserve their centre of mass momentum \cite{Regal04oor}.  While effective, this method relies on the interplay of expansion and pairing dynamics \cite{Matyjaskiewicz08pfc}, and, due to the necessity for expansion, is incompatible with obtaining local information.

An alternative signature of macroscopic order is the collective (Goldstone) mode \cite{Anderson58rpa}, a long-wavelength bosonic excitation with linear dispersion and gradient equal to the sound velocity \cite{Minguzzi01dsf,Ohashi03sac,Combescot06cmo}.  At large momenta, this mode evolves into a particle-like excitation with quadratic dispersion that, in two-component fermionic systems, physically represents the scattering of zero-momentum pairs from the condensate \cite{Combescot06msi,Veeravalli08bso,Son10sda,Hoinka13pdo}.  In this Letter, we study this mode in a trapped spin-balanced Fermi gas at unitarity using high momentum Bragg spectroscopy and find that it provides a dramatic signature for pair condensation that can be studied \emph{locally}.  

For 3D atomic gases, absorption imaging provides only a 2D projection of inhomogeneous atom clouds which integrates over regions with different density.  Thus, local information, such as the precise density or temperature at a phase boundary, is generally not accessible in a standard image.  Techniques such as the inverse Abel transform can reconstruct the local density, as was recently used for the measurement of the equation of state of the unitary Fermi gas \cite{Ku12rts}; however, one often wishes to know more than simply the density.  For example, measuring dynamical variables generally requires perturbing the system with a probe particle or photon which can destroy the (elliptic) cylindrical symmetry necessary for the inverse Abel transform \cite{Ku12rts}.  Recently, it was shown that probing a small region near the centre of an inhomogeneous cloud can provide a good representation of a homogeneous system \cite{Drake12doo,Sagi12mot}.  Here, we present an alternative scheme that does not require imaging of clouds with (elliptic) cylindrical symmetry and facilitates the measurement of dynamic variables including the dynamic spin susceptibility \cite{Hoinka12dsr}, density-density response \cite{Veeravalli08bso}, as well as Tan's universal contact \cite{Tan08eoa,Kuhnle10ubo,Sagi12mot}.  The method generalizes a scheme for obtaining the local pressure from the 1D line density \cite{Ho10otp,Nascimbene10ett} and shows that this conceptual approach is more powerful than previously realized.  

\begin{figure}[ht]
\begin{center}
\includegraphics [width=0.48\textwidth]{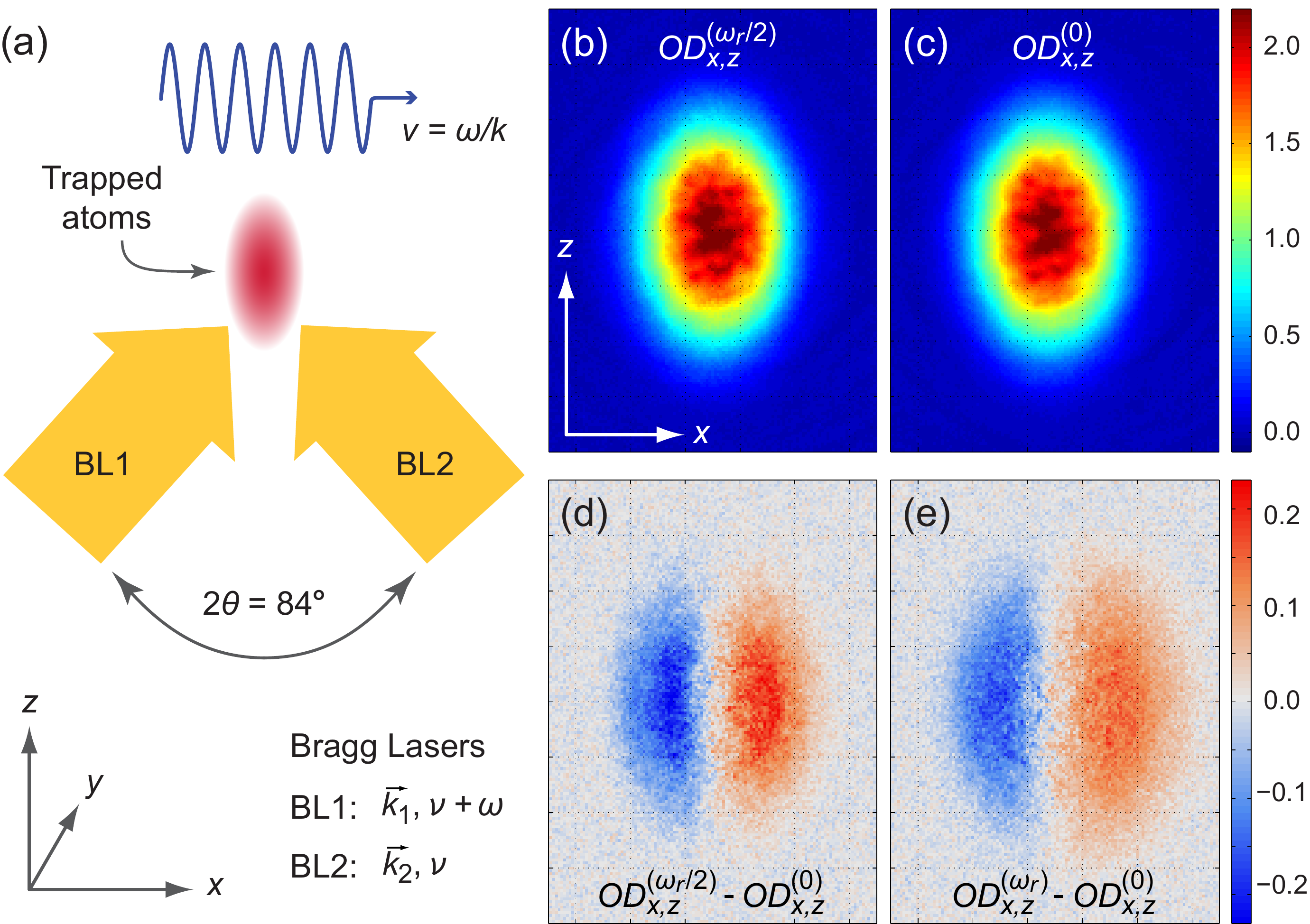} 
\caption{(a) Experimental arrangement for measuring the Bragg response.  Two laser beams with wavevectors $\vec{k}_1$ and $\vec{k}_2$, intersecting at an angle of $2 \theta = 84^{\circ}$, produce a moving interference pattern that illuminates a trapped atom cloud.  The long axis of the cloud is aligned along $z$ and Bragg scattering imparts momentum along $x$. (b) and (c), averaged optical density images, $OD_{ x,z }^{(\omega)}$, of an atom cloud following a relatively strong Bragg kick at a frequency of $\omega = \omega_r/2$ (b), and no Bragg kick at $\omega = 0$ (c).  These images appear nearly identical due to the short ($300 \, \mu$s) time of flight; however, subtracting them (d) shows that atoms have been displaced from left to right.  Another difference image is shown in (e) for two images obtained at Bragg frequencies of $\omega_r$ and 0.  The two difference images are themselves subtly different; the atom displacement for $\omega_r/2$ (d) is concentrated close to the center of the cloud while displacement for $\omega_r$ (e) is more diffuse with greater response from the wings of the cloud.  The viewing area of all images in (b) - (e) is $340 \, \mu\mathrm{m} \times 510 \, \mu\mathrm{m}$.}
\end{center}
\label{fig1}       
\end{figure}

Consider the measurement of the imaginary part of the dynamic susceptibility (density-density response) $\chi_{(k,\omega)}''$ using Bragg spectroscopy, Fig.~1(a), where $k$ is the Bragg wavevector and $\hbar \omega$ is the Bragg energy.  A bulk Bragg spectrum $\chi''^{\mathrm{B}}_{(k,\omega)}$, representing the density-averaged response of an inhomogeneous atom cloud, is obtained by illuminating the atoms with two Bragg lasers intersecting at an angle of $2 \theta = 84^{\circ}$ and measuring the total momentum imparted to the cloud as a function of $\omega$ \cite{Stenger99bso,Brunello01mtt}.  The atomic recoil frequency is defined as $\omega_r = \hbar k^2/(2m)$ where $m$ is the mass of a single atom.  Bragg spectroscopy has previously yielded the bulk dynamic and static structure factors of trapped Bose \cite{Stenger99bso,Steinhauer02eso} and Fermi gases \cite{Veeravalli08bso,Hoinka13pdo}, as well as Tan's universal contact parameter \cite{Tan08eoa,Hoinka13pdo}. 

In the experiments which follow we use an evaporatively cooled cloud containing a balanced mixture of approximately $N$/2 = 250,000 $^6$Li atoms in each of the lowest two spin states $| F = 1/2, m_F = \pm 1/2 \rangle$.  Atoms are confined in a highly harmonic hybrid optical-magnetic trap with frequencies of $(\omega_x, \omega_y, \omega_z) = 2 \pi \times (36.4, 250, 24.5)$ s$^{-1}$ at a magnetic field of 833$\,$G where the $s$-wave scattering length diverges (unitarity limit) \cite{Zurn13pco}.  A single mode 1064$\,$nm fiber laser that is spatially filtered before entering the vacuum cell produces the optical trap and magnetic confinement arises from a slight curvature of the (833$\,$G) magnetic field.  The bulk (trapped) Fermi energy is defined as $E_F^{\mathrm{B}} = k_B T_F^{\mathrm{B}} = (3N)^{1/3} \hbar \bar{\omega}_{ho}$ where $k_B$ is Boltzmann's constant and $\bar{\omega}_{ho} = (\omega_x \omega_y \omega_z)^{1/3}$ is the geometric mean confinement frequency.  After evaporative cooling, we typically produce clouds with temperatures of $0.08 \, T_F^{\mathrm{B}}$.  Higher temperatures are obtained by releasing the atoms from the optical trap for a variable time before recapturing and holding for a further 500$\,$ms for re-equilibration, or, by varying the end point of the evaporation.  Temperatures are determined by fitting the equation of state for the pressure of a unitary Fermi gas \cite{VanHoucke12fdv} to the line density of trapped clouds.  Bragg spectroscopy is performed by pulsing on the Bragg lasers for 100$\,\mu$s and measuring momentum transferred to the cloud from the resultant centre of mass displacement \cite{Veeravalli08bso,Hoinka13pdo}.  The Bragg lasers are detuned approximately 600$\,$MHz from the nearest atomic transition to probe the density-density response \cite{Hoinka12dsr}.

As the atom cloud is elongated along $z$, and the Bragg lasers transfer momentum to the atoms in the $x$-direction, it becomes possible to resolve the response from different $z$-positions along the cloud provided a short time of flight is used (compared to the timescale for dynamics along $z$).  Figure 1(b) and (c) show optical density images, $OD_{x,z}^{(\omega)}$, (averaged over 10 runs of the experiment under the same conditions) $300 \, \mu$s after a Bragg pulse was applied with Bragg frequencies of $\omega = \omega_r/2$ (b), and $\omega = 0$ (c, no Bragg kick), respectively.  While these images appear nearly identical, subtracting them (d) reveals that the Bragg pulse not only displaces atoms from left to right, but that the strongest response comes from the center of the cloud.  Furthermore, for different frequencies the $z$-dependence of the response changes.  Figure 1(e) shows a difference image for $\omega = \omega_r$ where the response is less intense but originates from a broader area of the cloud. 

To analyze these images we determine a $z$-dependent line response function, $\tilde{\chi}_{(k,\omega)}''(z)$, which quantifies the atom displacement as a function of $z$.  This is found by dividing the image into a series of horizontal strips (typically 10 to 30 $\mu$m wide) and evaluating the (left to right) centre of mass displacements within each strip.  $\tilde{\chi}_{(k,\omega)}''(z)$ is given by the density-weighted response function integrated over $x$ and $y$ \cite{suppmat}:
\begin{equation}
\tilde{\chi}_{(k,\omega)}''(z) = \frac{1}{\tilde{n}(z)} \int_{-\infty}^{\infty}{ \chi_{(k,\omega)}'' (\mu(\mathbf{r}), T) n( \mu(\mathbf{r}), T)  \mathrm{d}x \mathrm{d}y},
\label{eq1}  
\end{equation}
where $\tilde{n}(z) = \int{n( \mu(\mathbf{r}), T)  \mathrm{d}x \mathrm{d}y}$ is the (doubly-integrated) line density and $\chi_{(k,\omega)}''(\mu(\mathbf{r}), T)$ and $n(\mu(\mathbf{r}), T)$ are the local response and density of a cloud with chemical potential $\mu(\mathbf{r})$ and temperature $T$, respectively.  Equation (1) assumes the local density approximation (LDA) where $\mu(\mathbf{r}) = \mu_0 - V(\mathbf{r})$, $\mu_0$ is the chemical potential at the trap centre and $V(\mathbf{r})$ is the confining potential.  We expect the LDA to be valid for $\chi_{(k,\omega)}'' (\mu(\mathbf{r}), T)$ at large $k$, as the Bragg response is primarily determined by correlations on a length scale of $\lesssim k^{-1}$.  In our experiments, $k^{-1} = 80 \,$nm which is much smaller than the mean harmonic oscillator quantization length $l_{ho} = \sqrt{\hbar / (m \bar{\omega}_{ho})} = 5 \, \mu$m.  Thus, provided the atomic density also satisfies the LDA, (i.e. $\mu \gg \hbar \bar{\omega}_{ho}$) as has been validated experimentally \cite{Nascimbene10ett}, Eq.~(1) will be valid here.

For a gas confined in a harmonic potential, Eq.~(1) can be rewritten as an integral over the chemical potential using $\mathrm{d}x \mathrm{d}y = -2 \pi / (m \omega_x \omega_y)\mathrm{d}\mu$, where $\omega_x$ and $\omega_y$ are the trapping frequencies in the $x$ and $y$ directions, respectively \cite{Ho10otp}.  Making this substitution, differentiating with respect to $z$ and rearranging \cite{suppmat} we extract the \emph{local} homogeneous value of $\chi_{(k,\omega)}'' (\mu_z, T)$ along the axis of the trap,
\begin{equation}
\chi_{(k,\omega)}'' (\mu_z, T) = \frac{ \partial \left ( \tilde{\chi}_{(k,\omega)}'' (z) \, \tilde{n}(z) \right ) } { \partial \tilde{n}(z) },
\label{eq2}
\end{equation}
where $\mu_z = \mu_0 - V(0,0,z)$ . This simple relation connects the local response along the axis of the trap to the derivative of the line response multiplied by the line density.  We emphasize that this procedure is completely general and can be adapted to provide the local value of any quantity satisfying the LDA.  The images required for Eq~(2), (Fig.~1(d) and (e)), no longer satisfy the symmetry requirements for performing an inverse Abel transform.  The only requirement is that the cloud was initially confined in a harmonic potential.

\begin{figure}[ht]
\begin{center}
\includegraphics [width=0.48\textwidth]{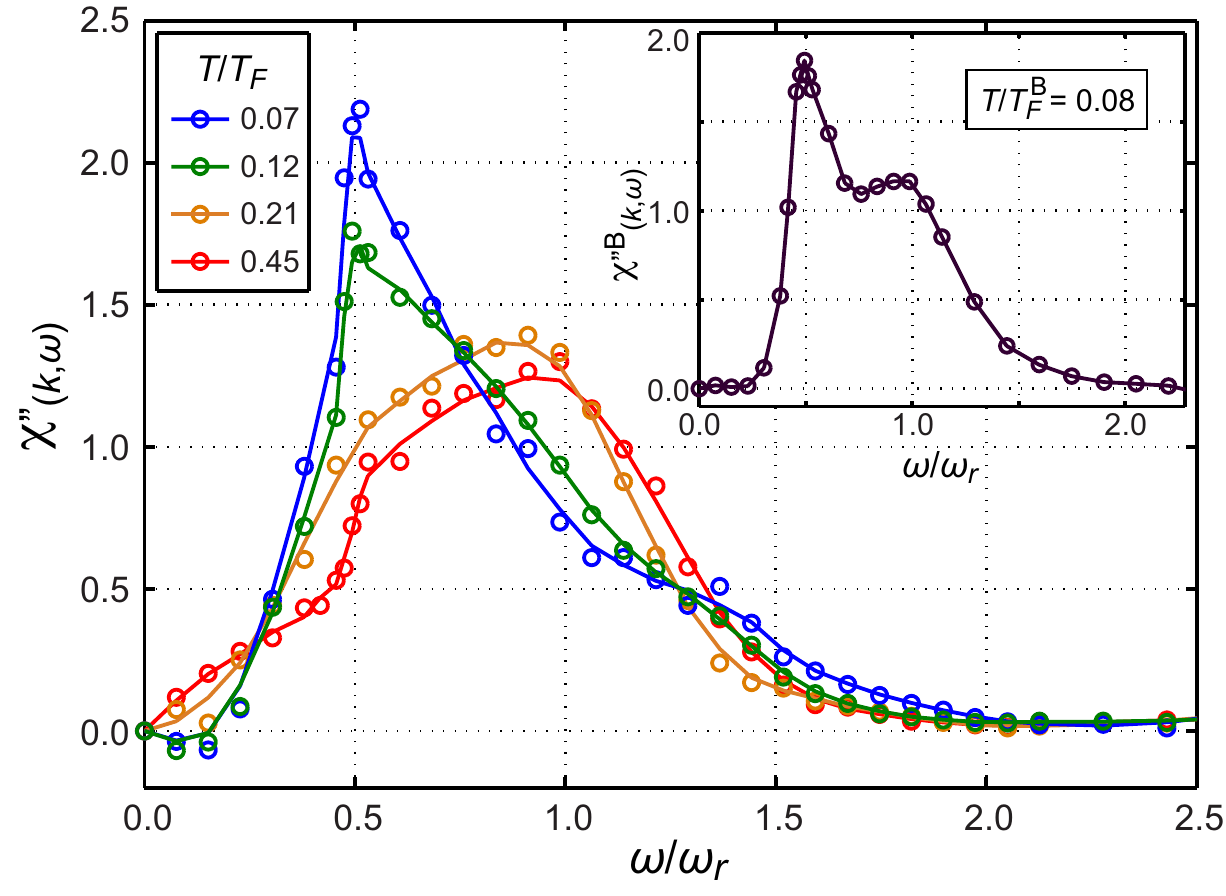} 
\caption{Local (homogeneous) Bragg spectra of a unitary Fermi gas measured below (blue, green) and above (orange, red) the superfluid transition temperature $T_c = 0.167 \, T_F$ \cite{Ku12rts}.   Local values of $T/T_F$ and $k/k_F$ contributing to each measurement are given in the text.  Solid lines are a guide to the eye.  Spectra below $T_c$ are dominated by a sharp feature at $\omega_r/2$ corresponding to pair scattering while the spectra above $T_c$ show a rounder response peaked just below $\omega_r$ corresponding to the continuum of single particle excitations.  Inset: Bulk response of a trapped unitary Fermi gas at a temperature of $T = 0.08 \, T_F^{\mathrm{B}}$ (from \cite{Hoinka13pdo}) which shows a weaker pairing signature at $\omega_r/2$, as well as a broader peak centered near $\omega_r$, due to the averaging over a range of relative temperatures and wavevectors in an inhomogeneous trapped cloud.}
\end{center}
\label{fig2}       
\end{figure}

We now proceed to the measurement of the homogeneous response $\chi_{(k,\omega)}''$.  At unitarity this will be a universal function of the relative temperature, $T/T_F$, and wavevector, $k/k_F$, where $T_F$ is the local Fermi temperature.  While $k$ and $T$ are uniform across the cloud, this method allows us to probe a range of $T/T_F$ and $k/k_F$ values simultaneously using a single cloud.  The local density sets the energy scale through the Fermi energy $E_F(\mathbf{r}) = k_B T_F(\mathbf{r}) = (\hbar^2/2m) (3 \pi^2 n(\mathbf{r}))^{2/3}$.  Similarly, the Fermi wavevector varies as $k_F(\mathbf{r}) = (3 \pi^2 n(\mathbf{r}))^{1/3}$.  The local response along the trap axis will therefore span a range of $T/T_F$ and $k/k_F$ as the density along the trap axis changes.  We find the local density $n(0,0,z)$ either from the derivative of the line density \cite{Ho10otp,Cheng07trf} or from the inverse Abel transform \cite{Ku12rts} of a trapped cloud before Bragg scattering. 

Figure~2 main panel shows the local (homogeneous) response functions, constructed using Eq.~(2), for atoms below (blue, green) and above (orange, red) the superfluid transition temperature $T_c = 0.167 \, T_F$ \cite{Ku12rts}. Local values of $T/T_F$ and $k/k_F$ are (0.07$\,^{+0.04}_{-0.03}$, 0.12$\,^{+0.04}_{-0.03}$, 0.21$\,^{+0.04}_{-0.03}$ and 0.45$\,^{+0.15}_{-0.10}$) and (3.8$\,^{+0.4}_{-0.3}$, 3.9$\,^{+0.5}_{-0.3}$, 4.3$\,^{+0.5}_{-0.3}$ and 5.3$\,^{+0.9}_{-0.5}$), for the blue, green, orange and red spectra, respectively.  Error bounds include the spread of momenta and temperatures arising from the range of densities contributing in each measurement.  Spectra were obtained using clouds with different initial temperatures so that the temperature dependence could be compared at similar $k/k_F$.  All spectra are averaged over a $\sim$ 30 $\mu$m region along $z$, centered approximately 0.2-0.3 of the cloud radius from the trap center, over which the local density varies by less than $15\%$, to improve signal-to-noise.  Each spectrum is normalized using the $f$-sum rule so that integration over all $\omega$ directly yields the homogeneous static structure factor \cite{Hoinka13pdo,Kuhnle10ubo}.  

Below $T_c$, the spectra are dominated by a sharp peak at $\omega_r/2$, corresponding to pair scattering, which tails off at higher frequencies, in good qualitative agreement with zero-temperature dynamical mean field theory~\cite{Combescot06msi}.  Above $T_c$, the response is more rounded and peaked closer to the atomic recoil frequency $\omega_r$ where single particle scattering dominates.  Pair scattering in the spectra below $T_c$ displays a very strong temperature dependence whereas above $T_c$, the spectra possess no sharp features and show a much weaker temperature dependence.  Also plotted (inset) is a bulk spectrum of a cold trapped gas which shows both pair and single particle peaks due to the averaging over different densities (from \cite{Hoinka13pdo}).
    
\begin{figure}[htpb]
\begin{center}
\includegraphics [width=0.48\textwidth]{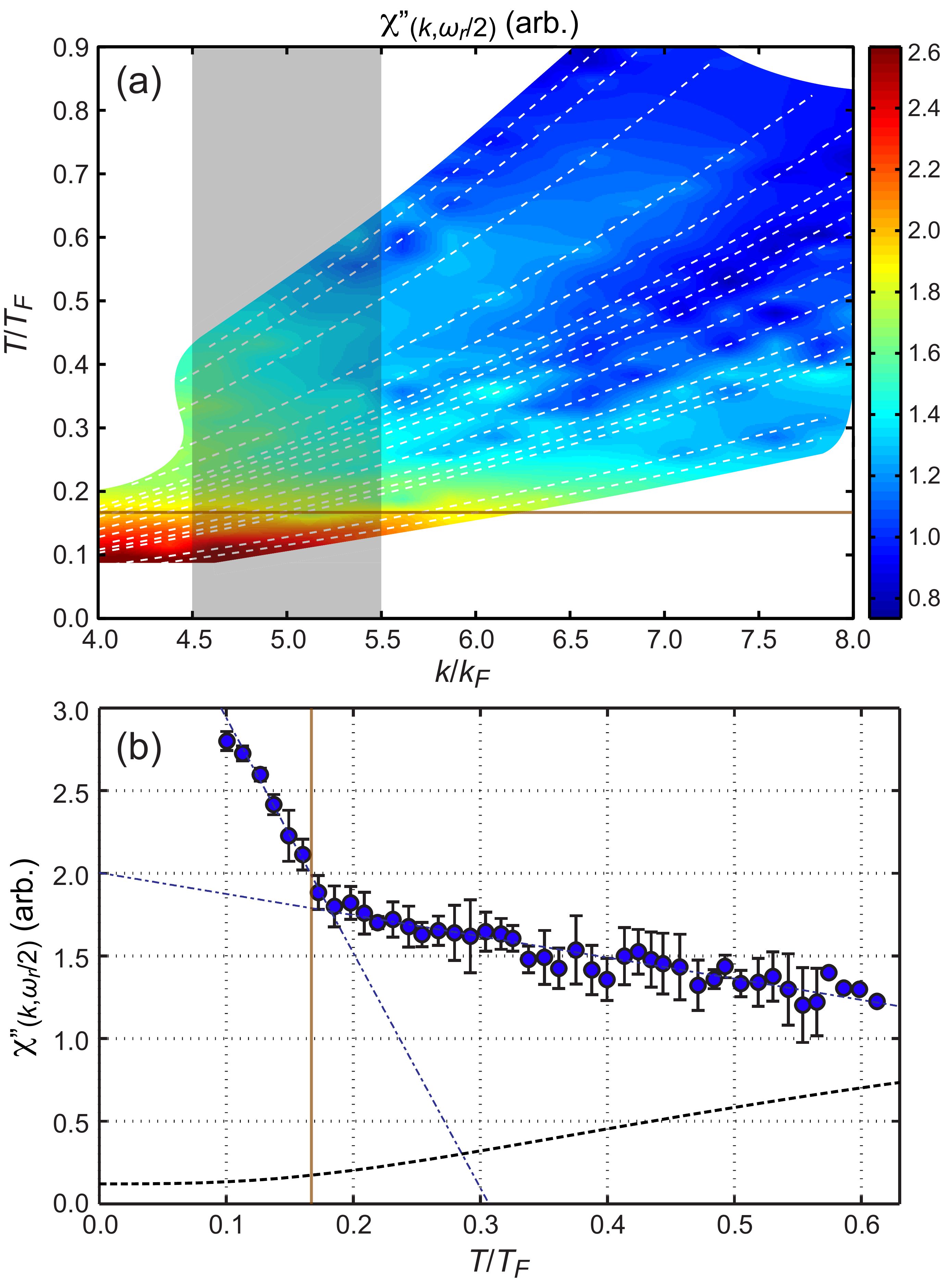} 
\caption{(a) False color image of the local (homogeneous) response $\chi_{(k,\omega_r/2)}''$ of a unitary Fermi gas as a function of the relative temperature $T/T_F$ and wavevector $k/k_F$.  The image was constructed by measuring the local response for a range of clouds prepared at different initial temperatures and binned according to temperature and wavevector.  White dashed lines show the temperatures and wavevectors spanned by individual (averaged) clouds.  A rapid increase in the response is observed for temperatures $T/T_F < 0.2$, due to the appearance of the pair condensate.  (b) Local response versus relative temperature for wavevectors in the range $4.5 < k/k_F < 5.5$, using data indicated by the grey shaded region in (a).  This smaller range of momenta shows the sudden increase in the response below $T_c = 0.167 \, T_F$ \cite{Ku12rts}, (brown lines in (a) and (b)).  Also shown (black dashed line) is the calculated response for an ideal gas at $\omega_r/2$ for $k = 5 \, k_F$ which shows the opposite temperature dependence to the data.  Blue dash-dotted lines are straight line fits to the data over the temperature ranges $0.11 < T/T_F < 0.17$ and $0.2 < T/T_F < 0.6$ used to estimate the critical temperature. }
\end{center}
\label{fig3}       
\end{figure}

Next we perform measurements of the Bragg response at a frequency of $\omega_r/2$ corresponding to the top of the pair peak.  For this we use a 200 $\mu$s Bragg pulse to increase signal-to-noise and improve spectral resolution.  Several clouds with bulk temperatures ranging from  $0.08 \, T_F^{\mathrm{B}}$ to $0.6 \, T_F^{\mathrm{B}}$ were used and the measured (local) centre of mass displacements were binned according to $T/T_F$ and $k/k_F$ to produce a false colour image of the local response, $\chi_{(k,\omega_r/2)}''$, Fig.~3(a).  White dashed lines show the range of temperatures and wavevectors spanned by individual clouds (averaged of 20 experimental runs) used to construct the image.  At high temperatures the response is relatively flat showing a weak dependence on $T/T_F$ and $k/k_F$; however, a sharp increase is observed below $T/T_F \sim 0.2$.  Examining the response over the smaller range of wavevectors $4.5 < k/k_F < 5.5$, indicated by the shaded region in Fig.~3(a), which includes data both above and below $T_c$, we can more clearly see the temperature dependence, Fig.~3(b).  At high temperatures ($> 0.2 \, T_F$), the response increases slowly with decreasing temperature.  However, a rapid increase occurs when the temperature drops below $0.18 \, T_F$.  Also shown is the calculated response of an ideal Fermi gas at $k = 5\, k_F$ (black dashed line).

Due to energy and momentum conservation at the Bragg condition, the sudden increase in the response at $\omega_r/2$ below $T_c$ signifies the accumulation of zero-momentum pairs in the condensate.  Despite the fact that the Bragg recoil energy is more than ten times larger than the pairing gap $\Delta$ ($\Delta = 0.44 \, E_F$ at unitarity \cite{Schirotzek08dot}), two atoms can still scatter as a pair provided any increase in their relative energy is less than $\Delta$.  At $k \sim 5 \, k_F$, this collective (paired) mode at unitarity lies within the continuum of single particle excitations \cite{Combescot06cmo}, yet remains highly visible at $~ \omega_r/2$ \cite{Combescot06msi}, so both pair and single particle scattering contribute significant weight to our measurements.  It is interesting to note that the sudden appearance of the pair peak does not coincide with a strong enhancement of the contact below $T_c$ \cite{Sagi12mot,Kuhnle11tdo}.  This highlights the difference between the dynamic response, which reveals the pairing peak, and the integrated (static) response used to obtain the contact \cite{Hoinka13pdo}.

Empirically, we find that the data in Fig.~3(b) in the vicinity $T_c$ is approximately linear with two different slopes above and below the transition temperature.  Fitting straight lines over the ranges $0.11 < T/T_F < 0.17$ and $0.2 < T/T_F < 0.6$ (blue dash-dotted lines), we estimate the critical temperature for pair condensation from the intercept to be $T_c = 0.18\,^{+0.03}_{-0.02} \, T_F$, in good agreement with the recent thermodynamic determination of $0.167 \,T_F$ (brown lines, Fig 3) \cite{Ku12rts}.  Our error bars include uncertainties arising from the finite time of flight and spatial averaging ($\sim 20 \, \mu$m) used in this measurement.  We note that even though the peak associated with pair condensation is visible in the bulk Bragg spectrum (Fig. 2, inset), we cannot use bulk measurements to determine $T_c$ as density averaging necessarily includes a large spread of different relative temperatures and wavevectors in the bulk response.  Instead of showing a sudden change at $T_c$, the bulk response at $\omega_r/2$ displays only a smooth and more gradual increase as the temperature is lowered.  

In the normal phase, above $T_c$, the measured response displays the opposite temperature dependence to an ideal gas.  This shows the build up of short-range pair correlations as the temperature is lowered; however, as both pair and single atom scattering are present, we cannot identify this as the scattering of non-condensed (bound) pairs \cite{Gaebler10oop}.  Local measurements of the dynamic spin susceptibility \cite{Hoinka12dsr} at $k \lesssim k_F$ could be used to clarify this issue of pseudogap pairing.  At higher temperatures the response should approach the ideal gas result which turns over near $2\,T_F$ and begins decreasing (not shown).   

In summary, we have shown that a scheme developed to measure the local pressure using a (non-uniform) harmonically trapped quantum gas \cite{Ho10otp} is more powerful than previously realized and is capable of yielding local parameters not accessible by conventional methods such as the inverse Abel transform.  We have used this technique to make the first measurements of the homogeneous density-density response function $\chi_{(k,\omega)}''$ of a Fermi gas at unitarity using Bragg spectroscopy.  Measuring the local response allows us to connect features in the Bragg spectrum with a specific local temperature, revealing a strong signature of pair condensation when the temperature drops below $T_c$.  

We thank H. Hui, M. Davis, P, Hannaford, R. Combescot and S. Stringari for helpful discussions.  C.J.V. acknowledges financial support from the Australian Research Council programs FT120100034 and DP130101807.



\newpage
\thispagestyle{empty}
\mbox{}

\newpage
 
\section{Supplemental material}




\noindent \textbf{Extracting homogeneous parameters from doubly-integrated (1D) line data}

\medskip

Consider the measurement of an arbitrary homogeneous parameter, $A(\mu, T)$, where $\mu$ is the chemical potential and $T$ is the temperature, in a harmonically trapped gas.  $A(\mu, T)$ could be any quantity such as a spectral function, determined by measuring the spin-flip rate in radio frequency (rf) spectroscopy, or even Tan's contact parameter which can be derived from the static structure factor or high frequency tails of rf or Bragg spectra.  We assume that $A(\mu(\mathbf{r}), T)$ satisfies the local density approximation (LDA), where $\mu(\mathbf{r}) = \mu_0 - V(\mathbf{r})$, $\mu_0$ is the chemical potential at the trap centre, $V(\mathbf{r}) = \frac{1}{2} m (\omega_x^2 x^2 + \omega_y^2 y^2 + \omega_z^2 z^2)$ is the confining potential and $m$ is the mass of the atoms.

Experiments with quantum gases typically confine atoms in a harmonic potential that is elongated along one direction, $z$.  Such a trap is well suited to this scheme as it is often possible to measure the one-dimensional parameter, $\tilde{A}(z)$, which is spatial resolved along the $z$-direction.  In analogy with Eq.~(1) in the main text, $\tilde{A}(z)$ will be given by
\begin{equation}
\tilde{A}(z) = \frac{1}{\tilde{n}(z)} \int_{-\infty}^{\infty}{ A(\mu(\mathbf{r}), T) n( \mu(\mathbf{r}), T)  \mathrm{d}x \mathrm{d}y},
\label{eq1}  
\end{equation}
where $\tilde{n}(z) = \int{n( \mu(\mathbf{r}), T)  \mathrm{d}x \mathrm{d}y}$ is the (doubly-integrated) line density.  In general, any 1D measurement will be density-weighted in this way reflecting the fact that the regions of the cloud with the largest number of atoms contribute the most signal to the measurement.  Making the substitution $\mathrm{d}x \mathrm{d}y = -2 \pi / (m \omega_x \omega_y)\mathrm{d}\mu$ \cite{Ho10otp} and multiplying both sides by $\tilde{n}(z)$ gives
\begin{equation}
\tilde{A}(z) \cdot \tilde{n}(z) = \frac{2 \pi}{m \omega_x \omega_y} \int^{\mu_z}_{-\infty}{ A(\mu(\mathbf{r}), T) n( \mu(\mathbf{r}), T)  \mathrm{d} \mu},
\label{eq2}  
\end{equation}
where $\mu_z \equiv \mu(0,0,z)$ is the chemical potential along the $z$-axis and $\mu \rightarrow -\infty$ when $(x, y) \rightarrow \infty$.  Next we differentiate both sides of Eq.~(2) with respect to $z$ which gives
\begin{equation}
\frac{ \mathrm{d} \left (\tilde{A}(z) \cdot \tilde{n}(z) \right )}{ \mathrm{d} z} = \frac{2 \pi}{m \omega_x \omega_y} \left [  A(\mu(\mathbf{r}), T) n( \mu(\mathbf{r}), T)  \frac{\mathrm{d} \mu}{\mathrm{d} z} \right ]^{\mu_z}_{-\infty}.
\label{eq3}  
\end{equation}
Differentiating extracts $A(\mu(\mathbf{r}), T) n( \mu(\mathbf{r}), T)  \frac{\mathrm{d} \mu}{\mathrm{d} z}$ at the limits of integration.  The density goes to zero as $\mu \rightarrow -\infty$ so only the $\mu_z$ term has nonzero weight.  In the case of harmonic trapping $\frac{\mathrm{d} \mu_z}{\mathrm{d} z} = - m \omega_z^2 z$, so that 
\begin{equation}
\frac{ \mathrm{d} \left (\tilde{A}(z) \cdot \tilde{n}(z) \right ) }{ \mathrm{d} z} = - \frac{2 \pi \omega_z^2 z}{ \omega_x \omega_y} A(\mu_z, T) n(\mu_z, T)
\label{eq4}  
\end{equation}
where $n(\mu_z, T)$ and $A(\mu_z, T)$ are the local (homogeneous) density and parameter of interest respectively.

As shown by Cheng and Yip \cite{Cheng07trf} and Ho and Zhou \cite{Ho10otp}, and, in analogy with the steps leading to Eq.~(4), the homogeneous density $n(\mu_z, T)$ can be found from the $z$-derivative of the line density $\tilde{n}(z)$
\begin{equation}
n(\mu_z, T) = -\frac{\omega_x \omega_y}{2 \pi \omega_z^2 z}  \frac{ \mathrm{d}\tilde{n}(z)  }{ \mathrm{d} z} 
\label{eq5}  
\end{equation}
which is equivalent to the Gibbs-Duhem equation.  Substituting (5) into (4) one finds
\begin{equation}
\frac{ \mathrm{d} \left (\tilde{A}(z) \cdot \tilde{n}(z) \right ) }{ \mathrm{d} z} = A(\mu_z, T) \frac{ \mathrm{d}\tilde{n}(z)  }{ \mathrm{d} z} 
\label{eq6}  
\end{equation}
which gives the general form of Eq.~(2) in the main text
\begin{equation}
A(\mu_z, T) = \frac{ \mathrm{d} \left (\tilde{A}(z) \cdot \tilde{n}(z) \right ) }{\mathrm{d}\tilde{n}(z)}.
\label{eq7}  
\end{equation}

Thus one can obtain the local value of any homogeneous parameter along the axis of a harmonic trap from 1D (doubly-integrated) line measurements for quantities satisfying the LDA.  We note that in practice this procedure for obtaining local parameters requires differentiating experimental data which means that even subtle differences in the raw (integrated) images, such as those seen in Fig.~1(d) and (e) in the main text, can lead to significant differences in the homogeneous parameters being measured.  This also means this procedure is quite sensitive to experimental noise so averaging several measurements is typically necessary.

\end{document}